# Fake it till you predict it: data augmentation strategies to detect initiation and termination of oncology treatment

Valentin POHYER[a,b,d], Elizabeth FABRE[a,c], Stéphane OUDARD[a,c], Laure FOURNIER[a,c,d], Bastien RANCE[a,b,d]
[a] Hôpital Européen Georges Pompidou, AP-HP, Paris, France
[b] Inserm, Centre de Recherche des Cordeliers, Université Sorbonne-Paris Cité, Paris, France
[c] Inserm, PARCC, Paris, France
[d] Université Paris Cité, Paris, France
ORCiD ID: Valentin Pohyer (0000-0003-0445-8388), Elizabeth FABRE (0000-0002-1244-873X), Stéphane OUDARD (0000-0003-0893-285X), Laure Fournier (0000-0002-1878-0290), Bastien Rance (0000-0003-4417-1197)

**Abstract.** At the hospital, the dispersion of information regarding anti-cancer treatment makes it difficult to extract. We proposed a solution capable of identifying dates, drugs and their temporal relationship within free-text oncology reports with very few manual annotations. We used pattern recognition for dates, dictionaries for drugs and transformer language models for the relationship, combined with a data augmentation strategy. Our models achieved good prediction F1-scores, reaching 0.872. The performance of models with data augmentation outperforms those of models without. By inferring such models, we can now identify and structure thousands of previously unavailable treatment events to better apprehend solutions and patient response.

## 1. Introduction

In the Electronic Health Records (EHRs) information is often dispersed and non-standard. For example, information regarding patients' treatment can be found in the Computerized Provider Order Entry (CPOE), or in free-text clinical reports - especially for treatment received outside the hospital. Specialized systems can also be used, this is the case for chemotherapy, often managed in a dedicated software. In the softwares and free-text clinical documents, various terminologies can be used: structured and standardized in CPOE, or completely unstructured using acronyms, abbreviations, and so forth, in free-texts.

This dispersion, and quasi absence of standardization of the information makes it difficult for researchers to grasp the complete picture of the care process of patients.

A wide body of literature has contributed to the structuration of drug information from text, but a challenge remains in the identification of initiation and termination of treatments especially in languages other than English.

In this study, we sought to develop a supervised Natural Language Processing (NLP) methodology capable of extracting the initiation and termination information regarding tumor treatment from oncology reports with a limited annotation workload.

## 2. Methods

We used the medkit [1] python library (a clinical NLP python toolkit) for all text processing tasks.

*Patient Cohort And Dataset.* We leveraged the Clinical Data Warehouse from the HEGP [2]. We collected all oncology reports from patients having billing codes corresponding to at least one of the following locations: primary tumor of the ovary (C56), primary tumor of the lung (C34) and primary tumor of the kidney (C64). A total of 9,506 eligible patients were selected, including 5,450 men (57.33%) and 4,056 women. The mean age of the cohort was 73 years (±14 SD).

*Identifying Temporal Relations Between Dates And Drugs.* In a nutshell, our method can be described as follows: we annotated sentences to detect occurrences of dates and drugs. We then determined the relationship between the date and drug among one of five categories: Initiation of treatment, Termination of treatment, Continuation of treatment and Absence of relationship. A category “Ambiguity” was also added to describe sentences in which human experts could not identify a clear relationship.

*Pre-processing Of The Oncology Reports.* Text reports were splitted by paragraphs, which were then splitted as sentences based on punctuation and structure.

*Detecting Dates And Drugs.* Dates were identified using medkit [1] and the EDS-NLP [3] set of rules using pattern recognition (regular expressions). For the identification of drugs we leveraged two French drug dictionaries: the OMEDIT [4] list of oral anti-cancer drugs and the French ATC terminologies (limited to L01-class treatment) [5]. We enriched the dictionaries with additional synonyms and protocols manually identified in texts. The full list of anti-cancer treatments can be found on the github repository dedicated to this paper [6]. We relied on the IAMsystem [7] to match the drugs in sentences. We used the Jaccard distance, a threshold set to 0.8 and a minimum number of characters of 8 as parameters of the drug matching algorithm.

*Manual Annotation Of The Dataset*. We extracted 1,235 sentences from the corpus. After automated annotation of drugs and dates, we manually annotated the temporal relations between all possible combinations of drugs and dates within sentences using one of the five categories listed above. For example, if a sentence contains four dates and three drugs, we generated a collection of 4 x 3 = 12 combinations. 200 sentences were set aside as a test set, the remainder for training.

*Data Augmentation Strategy.* To automatically generate new data using the existing sentences and annotations as references, we randomly selected an annotated sentence, containing a drug and a date, and replaced the real drug and date by a randomly selected drug and date (see Figure 1). This strategy enables the generation of billions of combinations of sentences containing drug, date and temporal annotation from our small training dataset.

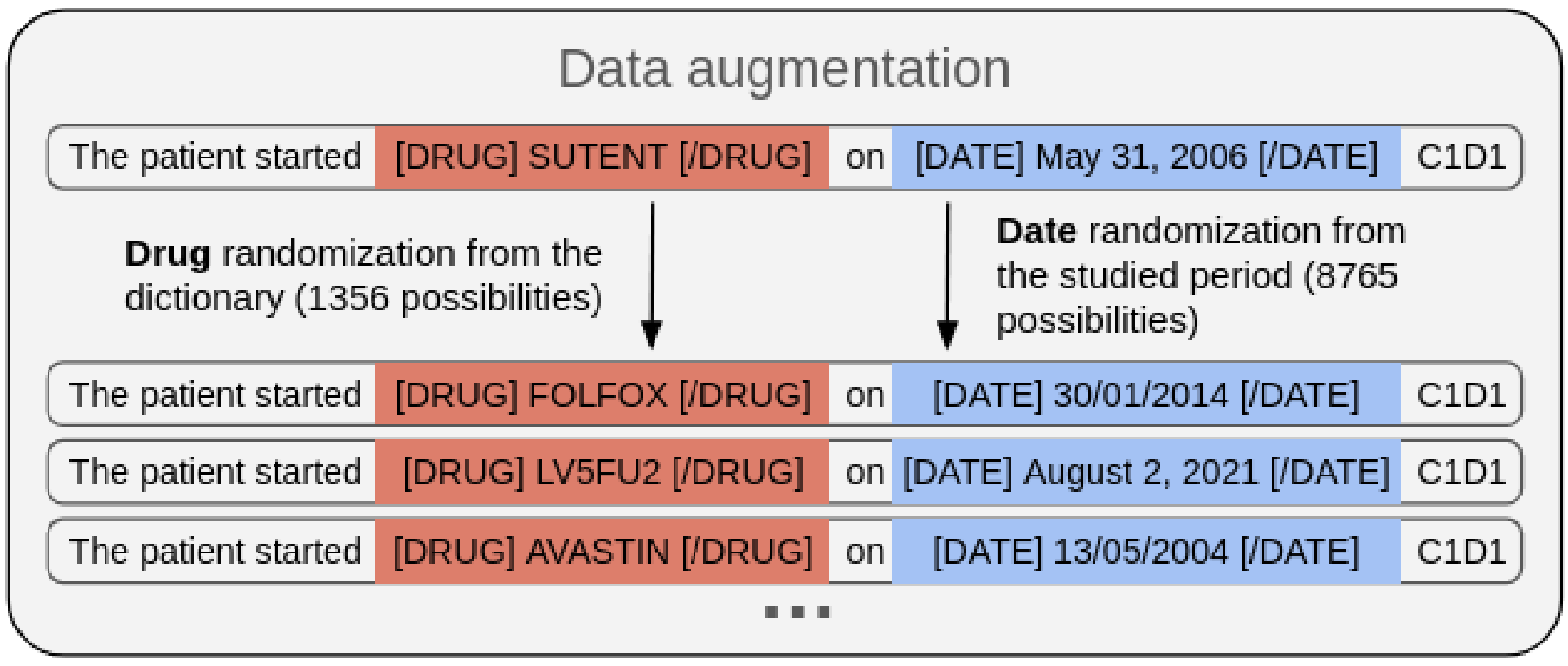

**Figure 1.** Illustration of the data augmentation process, generating factice sentences from the existing ones of the training dataset

*Predicting Temporal Relations Between Drug And Date Using A Language Model*. Using our original and augmented datasets, we trained models to classify the type of temporal relation (Initiation, Termination or Continuation of treatment, but also NoRelation and Ambiguity) associated with a date-drug combination. We leveraged four pre-trained French BERT models [8], namely: CamemBERT-bio-base, DrBERT-4GB-CP-PubMedBERT, DrBERT-7GB-Large and AliBERT-7GB. We tested the models using increasing size of datasets, first without data augmentation, then adding the automatically generated combinations (adding 3k, 9k, 21k, 45k and 93k generated sentences). The parameters used are available in Table 1.

*Parameters And Performance Evaluation*. Regardless of the volume used we performed a 70-30% split for training and evaluation. The maximum sentence processing size was set at 512 tokens. As for the hyperparameters, manual exploration enabled us to choose a batch size of 5, a learning rate of 1e-5 and a number of epochs of 6. For each of the five modalities, we measured precision, recall and F1-score. We used a bootstrap approach with 500 replications to obtain averaged evaluation results and 95% confidence intervals.

## 3. Results

*Patient Cohort and Dataset.* We extracted 200,942 oncology reports for the 9,506 eligible patients. These reports were splitted in 7,049,864 sentences, among which 317,510 sentences contained at least one date and one drug at the same time. These 317,510 sentences contained 875,793 date-drug combinations.

*Data Augmentation.* The augmented data set consisted of the same proportion of each modality as the training one, i.e. 23.0% treatment initiation, 14.6% treatment continuation, 10.5% treatment termination, 9.0% ambiguity and 42.9% no relationship.

*Predicting Temporal Relationships.* The results are summarized in Table 1. The results show a strong association between the performance and the size of the annotated dataset. Interestingly, the use of data augmentation always improved the f1-score in comparison to the baseline (defined as the training on the full manually annotated dataset). The performances of the other models are detailed on github [6].

*Inference.* We used the CamemBERT-bio-base model fine-tuned with a total of 96,235 annotations to annotate our complete dataset. We obtained the following predictions out of the 879,196 date-drug combinations: 196,143 Initiation of treatment, 82,520 Termination of treatment, 113,747 Continuation of treatment, 38,628 Ambiguity and 444,755 Absence of relationship.

**Table 1.** Detailed results obtained by the CamemBERT language model with different sizes of the training dataset. Numbers in parenthesis correspond to the gain on the baseline.

| Manual annot sentences | Date-drug annot | Generated annot | Total | | Initiation [95% CI] | Termination [95% CI] |
|---|---|---|---|---|---|---|
| 100 | 335 | | 335 | precision | .373 [.371 - .376] | .000 [.000 - .000] |
| | | | | recall | .839 [.836 - .842] | .000 [.000 - .000] |
| | | | | f1-score | .516 [.514 - .519] | .000 [.000 - .000] |
| 395 | 1230 | | 1230 | precision | .585 [.583 - .588] | .859 [.855 - .863] |
| | | | | recall | **.865 [.862 - .867]** | .557 [.552 - .562] |
| | | | | f1-score | .698 [.695 - .700] | .674 [.670 - .678] |
| 740 | 2406 | | 2406 | precision | **.737 [.734 - .739]** | .773 [.769 - .777] |
| | | | | recall | .851 [.848 - .853] | **.947 [.945 - .950]** |
| | | | | f1-score | **.789 [.787 - .791]** | **.850 [.848 - .853]** |
| 1035 | 3235 | **(baseline)** | 3235 | precision | .733 [.730 - .736] | **.912 [.908 - .916]** |
| | | | | recall | .851 [.848 - .853] | .544 [.539 - .549] |
| | | | | f1-score | .787 [.785 - .789] | .680 [.676 - .684] |
| 1035 | 3235 | 3000 | 6235 | precision | .802 [.800 - .805] | .863 [.859 - .866] |
| | | | | recall | .844 [.841 - .846] | .844 [.840 - .848] |
| | | | | f1-score | .822 [.820 - .824] (+0.035) | .852 [.850 - .855] (+0.172) |
| | | 9000 | 12235 | precision | .829 [.826 - .832] | .852 [.849 - .856] |
| | | | | recall | .825 [.823 - .828] | .845 [.842 - .849] |
| | | | | f1-score | .827 [.825 - .829] (+0.040) | .848 [.845 - .851] (+0.168) |
| | | 21000 | 24235 | precision | .831 [.828 - .834] | .856 [.853 - .860] |
| | | | | recall | .836 [.833 - .838] | **.870 [.867 - .874]** |
| | | | | f1-score | .833 [.831 - .835] (+0.046) | .863 [.860 - .865] (+0.183) |
| | | 45000 | 48235 | precision | **.871 [.869 - .874]** | .849 [.845 - .852] |
| | | | | recall | **.851 [.848 - .854]** | .816 [.812 - .820] |
| | | | | f1-score | **.861 [.859 - .862] (+0.074)** | .831 [.828 - .834] (+0.151) |
| | | 93000 | 96235 | precision | .848 [.845 - .851] | **.903 [.899 - .906]** |
| | | | | recall | .826 [.823 - .828] | .845 [.841 - .848] |
| | | | | f1-score | .836 [.834 - .838] (+0.049) | **.872 [.869 - .875] (+0.192)** |

## 4. Discussion

*Performances Of The Models.* Overall the performances on the augmented dataset surpasses all models based solely on manually annotated data. On manually annotated data, the performances improved with the size of the training set. Of the four models tested, each had a better asset for a given modality. To maximize the quality of inference, it would be advisable to predict with all of them and retain only the predictions made by the model best suited to each modality.

*Limits.* We chose to limit the size of the data-augmentation to a maximum of 93,000 sentences to keep the training costs to a reasonable level (less than 10 hours of training on a local NVIDIA A40 graphics processing unit). Our enhanced NLP strategy has enabled us to achieve good prediction scores, with F1-scores reaching 0.872 for Termination of treatment, improving by 19.2 points the performance compared to the manual annotation baseline. It would be possible to use Large Language Models to obtain results with a similar level of performance, but the energy and consequently environmental cost seems disproportionate (~100 inferences per second with the transformer model versus ~40s per inference with an average LLM).

*Data Sharing.* The sensitive nature of the data processed in this study does not allow us to make available either the date-drug temporal prediction obtained or the trained AI models.

*Impact Of This Study.* This study shows the feasibility of using data augmentation to improve the performance of machine learning models.

The performances of the model were good but could still be improved. Three strategies could be explored and combined. The manual annotation of a greater dataset, the usage of modality targeted model, and to make use of redundancies and complementarity between the different medical documents and data sources.

## 5. Conclusions

This study adds to the existing literature [10] by supporting the ability of NLP methods to automatically extract and integrate key information for the longitudinal follow-up of patients in hospitals. Until now, the lack of alignment between structured data and data present in free text ruled out the possibility of adopting such automatization methods. The use of such methods should be routinely programmed within hospital departments as a way to generate evidence from real-life data and support doctors in their practice, giving them more time for patient care.